\documentclass[twocolumn]{aastex631}

\usepackage[normalem]{ulem}
\usepackage{bm}
\usepackage{amsmath}

\graphicspath{{./}{figures/}}

\def\deg{\hbox{$^\circ$}}

\def\logcm{$\log c_{m}$\,}

\def\utw{\ensuremath{\smash{\rlap{\lower5pt\hbox{$\sim$}}}}}
\def\udtw{\ensuremath{\smash{\rlap{\lower6pt\hbox{$\approx$}}}}}

\begin{document}

\title{Clustering analysis of Ca {\sc i} 4227 line polarization using magnetohydrodynamic simulations of the solar atmosphere}

\author[0000-0001-5253-4213]{Harsh Mathur}
\affiliation{Indian Institute of Astrophysics, II Block, Koramangala, Bengaluru 560 034, India}

\author[0000-0003-4928-2021]{L. S. Anusha}
\affiliation{Indian Institute of Astrophysics, II Block, Koramangala, Bengaluru 560 034, India}

\author[0009-0003-8000-2179]{Devang Agnihotri}
\affiliation{Indian Institute of Astrophysics, II Block, Koramangala, Bengaluru 560 034, India}

\begin{abstract}
Ca {\sc i} 4227 \AA\, line is a strong resonance line formed in the Solar chromosphere.
At the limb, it produces the largest scattering polarization signal.
So far, modeling the linear polarization in this line has been limited to the use of one-dimensional semi-empirical models of the solar atmosphere. 
%
Using three-dimensional magnetohydrodynamical models of the solar atmosphere, in this paper, we perform 1.5D radiative transfer calculations to understand the formation of linear polarization profiles due to resonance scattering in this line at a near limb position.
We focus on studying the sensitivity of the resonance scattering polarization to the temperature and the density structures in the atmosphere.
We do not include the effects of magnetic and velocity fields in this study.
We use clustering analysis to identify linear polarization profiles with similar shape and group them accordingly for our study. 
We analyze the structure of the linear polarization profiles across 14 clusters, each representing different realizations of the solar atmosphere. Using source function ratio plots at various wing and core wavelength positions, we provide a qualitative explanation of linear polarization profiles in these clusters. 
\end{abstract}


\section{Introduction}
\label{sect:intro}

%
%
Observations of the solar chromosphere with high spatial, spectral, and temporal resolution along with high polarimetric sensitivity reveal complex and dynamic shapes of the polarization profiles of spectral lines.
Thus, understanding spectral line formation and their polarization properties is crucial, and three-dimensional (3D) radiative magnetohydrodynamic (MHD) simulations \citep[e.g.][]{2022A&A...664A..91P, 2011A&A...531A.154G, 2005A&A...429..335V, 1998ApJ...499..914S} are key to interpreting these observations.

With the advanced capabilities of the Daniel K. Inouye Solar Telescope \citep[DKIST;][]{2020SoPh..295..172R}, we anticipate high-resolution observations of scattering polarization in the Ca {\sc i} 4227 \AA\, line within quiet Sun regions.
The DKIST is expected to deliver spatial resolutions of around 0.1$\arcsec$ and temporal resolution of about 10 seconds.
Such precise measurements underscore the need for a robust theoretical grasp of polarization signals in Ca {\sc i} 4227 \AA\, spectral line.

Observations of the linear scattering polarization profile of the Ca {\sc i} 4227 \AA\, line exhibits a typical triple peak structure, which is known to be caused by the partial frequency redistribution (PFR) effects \citep[see, e.g.,][]{mf92, rene2005}.
The polarization signals at the line core are sensitive to magnetic fields in the lower solar chromosphere via the Hanle effect, while the polarization signals in the line wings produced by the PFR effects, are also sensitive to the photospheric magnetic fields through magneto-optical effects \citep{2018ApJ...854..150A}.
This spectral line is formed by a resonance transition between the ground level of neutral calcium, which has total angular momentum $J_l$ = 0 [$\mathrm{^1 S_0}$], and an upper level with $J_u$ = 1 [$\mathrm{^1 P^o_1}$].

The modeling of the Ca {\sc i} 4227 \AA\, line has so far been primarily done using 1D semi-empirical models assuming plane-parallel radiative transfer.
Studies based on semi-empirical models \citep{1985cdm..proc...67A, 1993ApJ...406..319F} have estimated the formation height of the core of this line to be between 0.9~Mm ($\mu$ = 1) and 1.1Mm ($\mu$ = 0.1) \citep[for eg.][]{rene2005, 2020A&A...641A..63C}.
Similar formation height of about 0.7~Mm at $\mu$ = 1 has also been reported by \citet{2024A&A...683A.207G} considering single vertical columns extracted from a 3D MHD simulation \citep[][]{2016A&A...585A...4C}.
\citet[][]{mf92} carried out a detailed analysis of the linear polarization in this line and the effects of PFR using VAL \citep[][]{val81} model atmosphere of the Sun.

The triple peak structure of the polarization profile in this line was qualitatively explained with the help of anisotropy factor and the unpolarized total source function gradients in \citet[][]{rene2005} who used the FALC model \citep[][]{fontenlaetal93}. 
\citet{LSA2011} modeled the forward scattering Hanle effect in the near disc center observations \citep{2011A&A...530L..13B} of the linear polarization of the Ca {\sc i} 4227 \AA\, line.
\citet{2014ApJ...793...42S} modeled the observed center-to-limb variation of the Ca {\sc i} 4227 \AA\, line.
\citet{2016ApJ...831L...5C, 2017ApJ...843...64C} highlighted the importance of accounting for the time evolution of the solar atmosphere to correctly model the scattering polarization signal of Ca {\sc i} 4227 \AA\, line before comparing with observations.
%

%
%
To interpret the forthcoming high-resolution observations from next-generation telescopes, such as DKIST, it is essential to model the Ca {\sc i} 4227 \AA\, line using realistic 
3D
MHD
simulations.
These simulations capture the dynamic and inhomogeneous behavior of the solar atmosphere.
Notably, \citet{2021ApJ...909..183J} investigated the effect of horizontal inhomogeneities and macroscopic velocity gradients in simulated disc center linear polarization.
To do this, they solved the full 3D non-LTE RT problem for polarized radiation, in the limit of complete frequency redistribution, using a publicly available Bifrost simulation \citep[][]{2016A&A...585A...4C}.
%
%
\citet{2024A&A...683A.207G} investigated the effects of angle-dependent PFR and bulk velocities on a few sample 1D models extracted from the Bifrost simulation.

In this paper, we make an attempt to understand the complex formation of the polarization profile of the Ca {\sc i} 4227 \AA\, line in a 3D MHD atmosphere from the publicly available Bifrost simulations \citep[][]{2016A&A...585A...4C} using 1.5D radiative transfer (RT) with PFR.

In particular we focus our attention on studying only the sensitivity of the resonance scattering polarization to the temperature and the density structures in the atmosphere.
Therefore we do not include the effects of magnetic and velocity fields in this paper.

A large number of 1D atmospheric structures can be obtained by piercing rays at different heliocentric angles ($\mu=\cos \theta$) through a 3D MHD cube atmosphere and interpolating the physical quantities along these rays (see e.g., \citet[][]{2021ApJS..255....3A} where a similar approach is followed). The interpolation along the ray uses the inhomogeneous, $x, y$ dependent physical parameters in the 3D MHD cube, thus covering a large range of densities and temperatures in the atmosphere.
This provides us an opportunity to understand the formation of linear polarization profiles in various atmospheric conditions.
By piercing rays at $\mu$ = 0.3 through the 3D MHD atmosphere we extract several 1D atmospheres.
Using angle-averaged PFR, we explain the formation of Ca {\sc i} 4227 \AA\,  polarization profiles emerging from these 1D atmospheres.
We employ the $k$-means clustering \citep{zbMATH03340881} technique to analyze the spectra of the large dataset resulting from the forward polarization synthesis of the 1D models extracted from the Bifrost simulation.
The effectiveness of the clustering techniques in the forward modeling context has been demonstrated successfully to correlate the shapes of spectral line intensities with the temperature structure of the atmospheric columns \citep{2023A&A...675A.130M}.

In Section~\ref{sect:theory}, we outline the underlying theory of polarized line formation that we used for the computations carried out in this paper.
In Section~\ref{sect:methods}, we describe the numerical method and computational details.
The clustering on the dataset is described in section~\ref{sect:kmean}.
The results of our computations are discussed in Section~\ref{sect:results}.
Finally, conclusions are presented in Section~\ref{sect:conclusions}.

\section{THE RADIATIVE TRANSFER FORMULATION}
\label{sect:theory}

\begin{figure*}[htbp]
    \centering
    \includegraphics{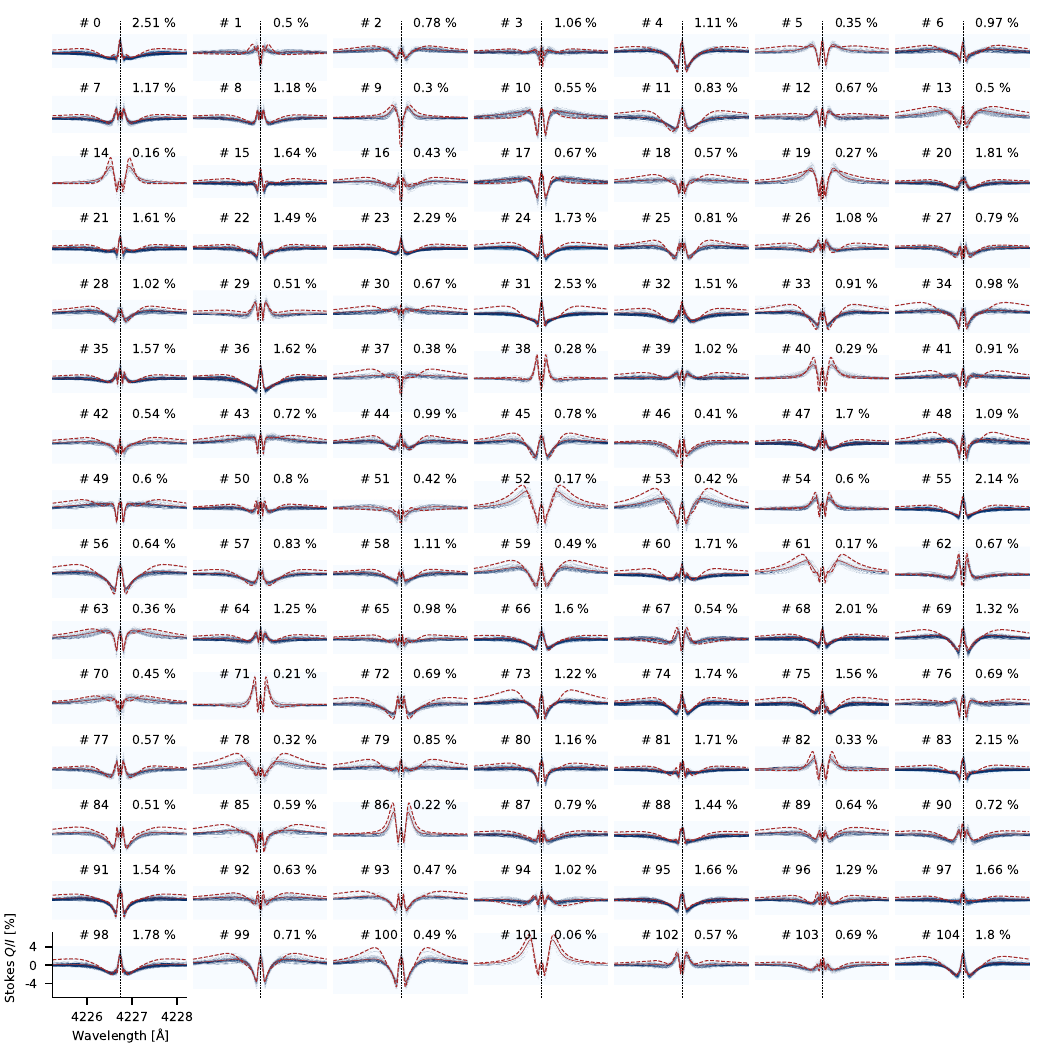}
    \caption{$k$-means clusters of Ca {\sc i} 4227 \AA\ synthesized spectra, using 105 clusters.  The density (darker, meaning a higher concentration of spectra) corresponding to each Ca {\sc i} 4227 \AA\ RP shows the distribution of profiles over the entire synthesized spectra. The solid red line denotes the average of all line profiles belonging to each cluster, i.e. the RP. The fraction of all profiles belonging to each cluster is indicated as a percentage next to the cluster number. The dashed red line denotes the spectral profile farthest (using $\chi^2$) from the red line. The black vertical line denotes the position of the nominal line center.}
    \label{fig:rps}
\end{figure*}

The theory that we use in this paper is already presented in several papers \citep[see][and the references therein]{LSA2011}.
Here, we summarize the necessary details for completeness.

\subsection{Polarized radiative transfer with resonance scattering}
\subsubsection{Stokes parameter formulation}

The Stokes vector RT equation for a two-level atom with unpolarised ground level in a 1D planar,  nonmagnetic medium without bulk velocities can be written as

\begin{equation}
\begin{split}
\mu\frac{\partial{\textrm{\bf{I}}}(\lambda,\boldsymbol{\Omega},z)}{\partial z} = -[\kappa_{l}(z)\phi(\lambda,z)+\kappa_{c}(\lambda,z)+\sigma_{c}(\lambda,z)]\\
\times [\textrm{\bf{I}}(\lambda,\boldsymbol{\Omega},z)-\textrm{\bf{S}}(\lambda,\boldsymbol{\Omega},z)],
\end{split}
\end{equation}

where $\textrm{\bf{I}} = (I,Q,U)^{T}$ is the Stokes vector and $\textrm{\bf{S}}= (S_{I},S_{Q},S_{U})^{T}$ the source vector.
Here, $\lambda$ is the wavelength, $z$ is the vertical height, and $\Omega=(\theta,\varphi)$ defines the direction of the ray where $\theta$ is the polar angle while $\varphi$ is the azimuth angle.
$\phi(\lambda,z)$ is the Voigt profile with damping parameter $a = {\Gamma_{\textrm{total}}}/{4\pi\delta \nu_{D}}$.
Here

\begin{align}
\Gamma_{\textrm{total}}= \Gamma_R+\Gamma_E+\Gamma_I.
\end{align}

The $\Gamma_R$ is the radiative de-excitation rate.
For Ca {\sc i} 4227 \AA\, line $\Gamma_R = 2.18\times 10^{8} s^{-1}. \Gamma_E$ is added FWHM of van der Waals broadening (elastic collision with neutral hydrogen) and Stark broadening (due to interaction with free electrons).
$\Gamma_I$ is the FWHM for inelastic processes which involves collisional de-excitation by protons and electrons and collisional ionization by electrons, and charge exchange processes \citep[see][]{2001ApJ...557..389U}.
$\Delta\nu_D$ = $\sqrt{\frac{2k_BT}{M_a}+v_{\textrm{turb}}^2}\frac{1}{\lambda_0}$ is the Doppler width, where $M_a$ is the mass of the atom, $v_{\textrm{turb}}$ is the micro-turbulent velocity (taken as 1 km/s), and $\lambda_0$ is the line center wavelength. $\kappa_l$ is the wavelength-integrated line absorption coefficient, while $\sigma_c$ and $\kappa_c$ are continuum scattering and absorption coefficients, respectively.
The total opacity coefficient for Ca {\sc i} 4227 \AA\, line and the continuum is $\kappa_{\textrm{total}} = \kappa_l(z)\phi(\lambda,z) + \sigma_c(\lambda,z)+\kappa_c(\lambda,z)$.
In a two-level with an unpolarized ground level, the source vector is defined as

\begin{eqnarray}
&&\textrm{\bf{S}}(\lambda,\boldsymbol{\Omega},z) = \frac{\kappa_l(z) \phi(\lambda,z)\textrm{\bf{S}}_l(\lambda,\boldsymbol{\Omega},z)}{\kappa_{\textrm{total}}(\lambda,z)} \\ \nonumber
&& +\frac{\sigma_c(\lambda,z)\textrm{\bf{S}}_c(\lambda,\boldsymbol{\Omega},z)+\kappa_c(\lambda,z)B_{\lambda}(z)\textrm{\bf{U}}}{\kappa_{\textrm{total}}(\lambda,z)}.
\end{eqnarray}

Here, $\textrm{\bf{U}}$ = $(1,0,0)^T$ and $B_{\lambda}$ is the Planck function at the line center.
Individual expression for line and continuum source vectors can be written as

\begin{equation}
\begin{split}
\textrm{\bf{S}}_l(\lambda,\boldsymbol{\Omega},z)=\epsilon B_{\lambda}(z)\textrm{\bf{U}}+\int_{-\infty}^{+\infty}\oint\frac{\hat{R}(\lambda,\lambda',\boldsymbol{\Omega},\boldsymbol{\Omega'},z)}{\phi(\lambda,z)}\\
\times \textrm{\bf{I}}(\lambda',\boldsymbol{\Omega'},z)\frac{d\boldsymbol{\Omega'}d\lambda'}{4\pi}.
\end{split}
\end{equation}

Here, $\hat{R}$ is the resonance scattering PFR matrix \citep[approximation III of][]{vb97a,vb97b} which uses angle-averaged PFR functions. The thermalization parameter $\epsilon$ = ${\Gamma_I}/{\Gamma_R+\Gamma_I}$.
The continuum source vector is

\begin{equation}
\textrm{\bf{S}}_c(\lambda,\boldsymbol{\Omega},z)=\oint \hat{P}(\boldsymbol{\Omega},\boldsymbol{\Omega'})\textrm{\bf{I}}(\lambda,\boldsymbol{\Omega'},z)\frac{d\boldsymbol{\Omega'}}{4\pi}.
\end{equation}

$\hat{P}$ is the Rayleigh scattering phase matrix as frequency coherence is assumed for the continuum.
Also, primed quantities denote incoming photons, while unprimed ones are for outgoing photons (after scattering).

\subsubsection{Spherical Irreducible Tensor Decomposition}

\begin{figure*}[htbp]
    \centering
    \includegraphics[width=\textwidth]{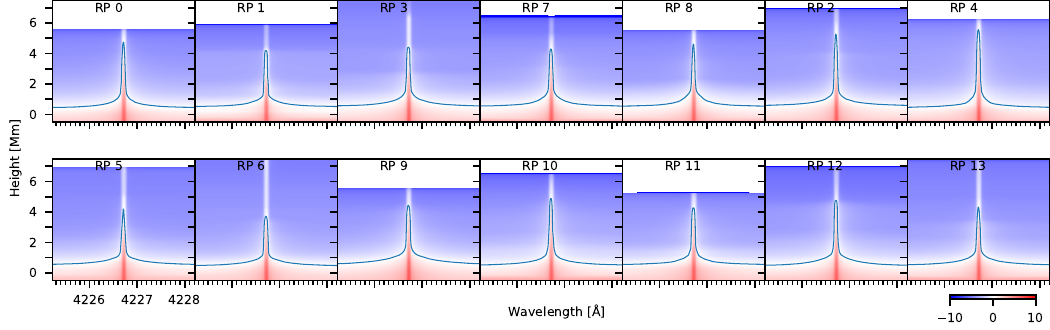}
    \caption{The optical depth, $\log (\tau_{\lambda}/\mu )$, is shown as a function of height, and wavelengths for a LOS with $\mu$ = 0.3. The $\log (\tau_{\lambda}/\mu)$ = 0 surface (blue) is overplotted for comparison. We remind the reader that the atmosphere is cropped based on the column mass scale, leading to varying height scales for different rays due to differences in their corresponding density structures.}
    \label{fig:tau}
\end{figure*}

The vectors $\textrm{\bf{I}}$ and $\textrm{\bf{S}}$ can be represented in terms of spherical irreducible tensors defined in \citet[][]{ll04}. The irreducible polarization and source vectors are denoted as ${\bm{\mathcal{I}}}$ and ${\bm{\mathcal{S}}}$ whose elements are $I^{K}_{Q}$ and $S^{K}_{Q}$, introduced by \citet[][]{hf07} \citep[see also][]{hf2022}, with $K=0,2$ and $Q$ $\in$ $[-K, +K]$.  
The advantage of doing this is if we construct source and Stokes vector in this irreducible form, then 
${\bm{\mathcal{S}}}$
becomes independent of the beam direction, and the ${\bm{\mathcal{I}}}$ becomes independent of azimuthal angle $\varphi$ for 1D medium.
The RT equation in this form for ${\bm{\mathcal{I}}}$ can be written as

\begin{equation}
\mu\frac{\partial{\bm{\mathcal{{I}}}}(\lambda,\mu,z)}{\partial z}  = -\kappa_{\textrm{total}}(\lambda,z)[{\bm{\mathcal{I}}}(\lambda,\mu,z)-{\bm{\mathcal{S}}}
(\lambda,z)].
\label{rte-ikq}
\end{equation}

Again, in a two-level model atom with an unpolarised ground state, the total irreducible 
source vector ${\bm{\mathcal{S}}}$ is defined as

\begin{equation}
\begin{split}
{\bm{\mathcal{S}}}(\lambda,z) = \frac{\kappa_l(z)\phi(\lambda,z){\bm{\mathcal{S}}}_l(\lambda,z)}{\kappa_{\textrm{total}}(\lambda,z)}\\
+\frac{\sigma_c(\lambda,z){\bm{\mathcal{S}}}_c(\lambda,z)+\kappa_c(\lambda,z)B_{\lambda}(z){\bm{\mathcal{U}}}}{\kappa_{\textrm{total}}(\lambda,z)}.
\end{split}
\end{equation}

Here, ${\bm{\mathcal{U}}}$ = $(1,0,0,0,0,0)^T$.
The irreducible line source vector is

\begin{eqnarray}
&& {\bm{\mathcal{S}}}_l(\lambda,z)=\epsilon B_{\lambda}(z){\bm{\mathcal{U}}} \nonumber \\
&&+\frac{1}{2}\int_{-\infty}^{+\infty}\int_{-1}^{+1}\frac{\hat{\mathcal{R}}(\lambda,\lambda',z)}{\phi(\lambda,z)}
\hat{\bm{\Psi}}(\mu')
\times {\bm{\mathcal{I}}}(\lambda',\mu',z)d\lambda'd\mu'.\nonumber \\
\label{line-skq}
\end{eqnarray}
$\hat{\mathcal{R}}$ is the angle-averaged PFR matrix for resonance scattering in the irreducible basis \citep[see e.g.,][]{LSA3D2011},
while ${\hat{\bm{{\Psi}}}}$ is the Rayleigh scattering phase matrix in the irreducible basis \citep[see e.g.,][]{hf07}.

Now, the continuum scattering source vector in irreducible basis is

\begin{equation}
{\bm{\mathcal{S}}}_c(\lambda,z) = \frac{1}{2}\int_{-1}^{+1}{\hat{{\bm{{\Psi}}}}(\mu'){\bm{\mathcal{I}}}}(\lambda,\mu',z)d\mu'.
\end{equation}

%
We define total optical depth scale as $d\tau_{\lambda}$ = $- \kappa_{\textrm{total}}(\lambda,z)dz$ and represent the total optical depth at the bottom of the atmosphere at a given wavelength $\lambda$ as $T_{\lambda}$.
General solution of Equation~(\ref{rte-ikq}) for $\mu>0$ can be written as

\begin{equation}
\begin{split}
{\bm{\mathcal{I}}}(\lambda,\mu,\tau_{\lambda}) = {\bm{\mathcal{I}}}_0(\lambda,\mu,T_\lambda)\exp\left(-\frac{T_\lambda-\tau_\lambda}{\mu}\right)\\
+ \int_{\tau_\lambda}^{T_\lambda}\exp\left(-\frac{\tau'_\lambda-\tau_\lambda}{\mu}\right){\bm{\mathcal{S}}}(\lambda,\tau'_\lambda)\frac{d\tau'_\lambda}{\mu},
\end{split}
\label{ikq-mup}
\end{equation}

and for $\mu<0$ the solution is

\begin{equation}
\begin{split}
{\bm{\mathcal{I}}}(\lambda,\mu,\tau_{\lambda}) = {\bm{\mathcal{I}}}_0(\lambda,\mu,0)\exp\left(-\frac{\tau_\lambda}{\mu}\right)\\
-\int_{0}^{\tau_\lambda}\exp\left(-\frac{\tau'_\lambda-\tau_\lambda}{\mu}\right){\bm{\mathcal{S}}}(\lambda,\tau'_\lambda)\frac{d\tau'_\lambda}{\mu}.
\end{split}
\end{equation}

It is assumed here that no radiation is coming into the medium from the upper boundary while at the bottom of the medium, LTE prevails; thus, we have ${\bm{\mathcal{I}}}_0$=0 for $\tau_\lambda$ = 0 and $\mu<0$.
While at $\tau_\lambda$ = $T_\lambda$ and $\mu>0$,  ${\bm{\mathcal{I}}}_0(\lambda,\mu,T_\lambda)$ = $(B_\lambda(T_\lambda),0,0,0,0,0)^T$.

\section{Modeling of the line polarization} 
\label{sect:methods}

To understand the formation of polarization profiles we carried out 1.5 polarized RT \citep[see, e.g.,][]{2021ApJS..255....3A} over hundreds of rays piercing through 3D MHD cube atmosphere.
These rays represent numerous manifestations of the dynamic solar atmosphere and, therefore, cover a large range of densities and temperatures in the atmosphere. 
This atmosphere is obtained from a publicly available enhanced network simulation \citep{2016A&A...585A...4C}, snapshot number 385, computed using the Bifrost Code \citep{2011A&A...531A.154G}.
%
%
For this work, we focus on resonance scattering polarization. Magnetic and velocity fields were not included.
The computations were performed for a heliocentric observing angle of $\theta_{LOS}$ = $72.54^{\circ}$ ($\mu$ = 0.3).
We interpolated the physical parameters along the slanted observing line-of-sight (LOS) in the simulation cube on a regularly spaced ($\Delta z$ = 20~km) Cartesian grid.
Thus the slanted ($\mu=0.3$) LOS rays in the original Bifrost box become the vertical directions in the new 1D models, and then we consider the radiation emergent at $\mu=0.3$ from the new models for our polarization studies.
Further, to reduce the computation time, we have only synthesized the spectral profiles for every 4th ray (pixel) in each direction ($x$ and $y$).
We calculated the $Q/I$ profiles for a total of 126$\times$126 = 15876 rays.

The calculations were performed in multiple steps, with the output of one step as input for the subsequent step.
Initially, we have utilized the RH \citep[][]{2001ApJ...557..389U} code to calculate the column mass ($c_m$) scale, and the atmosphere is reinterpolated and cropped between the \logcm{} = [$-$5, $+$1.5].
Further, using the reinterpolated atmosphere through the RH code, we calculated following quantities, continuum opacity, emissivity and scattering, line opacity and emissivity, mean intensity, collisional rates, and damping coefficient.
We have used the POLY \citep[see, e.g.,][]{rene2005,LSA2011} code, using the quantities calculated in the previous step to calculate the scattering polarization profiles, assuming that polarization does not affect the above-calculated quantities.

\section{Clustering of the $Q/I$ profiles}
\label{sect:kmean}

\begin{figure}
    \centering
    \includegraphics[width=0.5\textwidth]{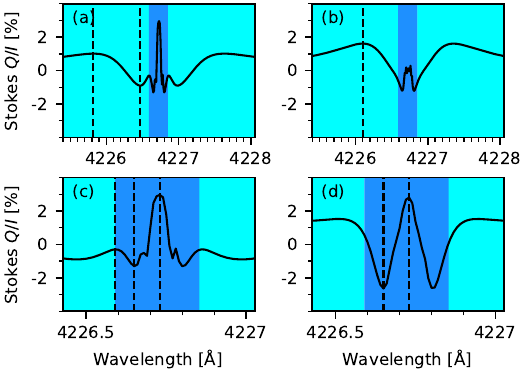}
    \caption{Examples of Stokes~$Q/I$ profiles are shown,  classified according to their shape in the core and the wings (see text). In the top panels we show the wing region, while in the bottom panels we show the core region with a primary core extremum at the line center. Panel (a) illustrates a profile with two wing extrema, while panel (b) depicts one with a single wing extremum.
     Panel (c)  presents a profile with a secondary core extremum and a core minimum, while panel (d) presents a profile with a core minimum without a secondary core extremum. The vertical lines mark the wavelength position of the extrema. The light and dark blue shaded colors show the wing and core regions, respectively.}
    \label{fig:classification}
\end{figure}

As stated in section~\ref{sect:methods}, $Q/I$ profiles are synthesized for a total of 15876 rays.
To interpret large datasets, algorithmic methods such as $k$-means clustering are routinely utilized in solar physics, reducing dimensionality, thereby facilitating efficient, qualitative and statistical analysis of millions of data points \citep{2018ApJ...861...62P, 2022A&A...664A..72J, 2022A&A...668A.153M, 2024A&A...682A..11M}.
We have partitioned the $Q/I$ profiles into 105 clusters using the $k$-means clustering technique, as explained in appendix~\ref{sect:appkmeans}.
Each $k$-cluster is represented by the mean of all the $Q/I$ profiles belonging to that cluster, which we refer to as the Representative Profile (RP).
In addition, for completeness, we also present here the results for the Stokes~$Q/I$ profile, which is farthest from the mean $Q/I$ profile, the RP.
Though the total number of clusters computed was 105, we have analyzed only 14 clusters.
This is because we found that the rest of the clusters have RP belonging to one of the shapes of the selected 14 clusters, with only slightly differing amplitude.
%
%
Using this clustering approach, we have made sure that all possible shapes of the $Q/I$ profiles have been taken into account for our analysis..

We show the results of the clustering of the synthesized spectra in Figure~\ref{fig:rps}.
For the most part, the profiles inside clusters are tightly constrained, as can be inferred from the dark-density plots.
The shapes of the $Q/I$ spectra are also not too dissimilar among different clusters.
The maximum amplitude observed in the synthesized  $Q/I$ profiles is approximately 6\%.
Although $k$-means clustering was performed with $k$ = 105, our analysis of cluster shapes suggests that the first 14 clusters adequately represent the main profile shapes and their varying amplitudes.
Therefore, subsequent analyses have been confined to these first 14 clusters.
The rest of the clusters are described in the appendix~\ref{sect:charac}.

\section{Results and discussions} 
\label{sect:results}
Qualitatively, the behavior of the  anisotropy factor ($J^2_0/J^0_0$) and the resulting $Q/I$ profiles are already explained in detail in \citet{rene2005} who used one semi-empirical atmosphere FALC. 
However, the shapes and magnitudes of the $Q/I$ profiles over numerous atmospheric models that we present here are very different when compared to the $Q/I$ profiles generated using the FALC atmosphere. 

To understand the formation of $Q/I$ we now analyze the polarization profiles based on source function ratios excluding the thermal source for the unpolarized component, rather than the standard anisotropy factor used in \citet{rene2005} as explained below. We define the irreducible source vector ${\bm{\mathcal{\tilde{S}}}}$ for 2-level atom system with PFR as 
 \begin{eqnarray}
&& {\bm{\mathcal{\tilde{S}}}}(\lambda,z)= \nonumber \\
&&+\frac{1}{2}\int_{-\infty}^{+\infty}\int_{-1}^{+1}\frac{\hat{\mathcal{R}}(\lambda,\lambda',z)}{\phi(\lambda,z)}
\hat{\bm{\Psi}}(\mu')
\times {\bm{\mathcal{I}}}(\lambda',\mu',z)d\lambda'd\mu',\nonumber \\
\label{line-jkq}
\end{eqnarray}
 
whose elements are denoted as $\tilde{S}^{K}_{Q}$. The first element  $\tilde{S}^0_0$ is the unpolarized line source component without the thermal source contribution and the second element $\tilde{S}^2_0$ represents the linearly polarized component of the line source vector (see Equation~\ref{line-skq}). 
In the absence of magnetic fields other elements of the vector vanish. 

We use the source function ratio  $\tilde{S}^2_0/\tilde{S}^0_0$ for our analysis, in which the elements 
$\tilde{S}^2_0$ and $\tilde{S}^0_0$ are respectively the elements $J^2_0$ and $J^0_0$ of the standard anisotropy factor $J^2_0/J^0_0$ \citep[][]{2001ASPC..236..161T}, multiplied by the term ${\mathcal{R}}(\lambda,\lambda',z)/\phi(\lambda,z)$ and integrated over wavelength.

We recall here that $\tau_{\lambda}$ is defined along the vertical (in the new models), so that $\tau_{\lambda}/\mu$ represents the optical depth along the considered LOS.
Further, due to the differences in density distribution in FALC and 3D MHD simulations of the solar chromosphere, we obtain a different distribution of optical depths ($\tau_{\lambda}/\mu$) and, hence, different heights of line formation.
This also affects the region where linear polarization is formed in these atmospheres.

Figure~\ref{fig:tau} displays a two-dimensional image of the $\log (\tau_{\lambda}/\mu)$ variation along wavelength and height for slanted geometry, $\mu$ = 0.3 for all the chosen clusters.
Continuum is formed at the base of the photosphere (height 0~Mm), and then the height at which $\log (\tau_{\lambda}/\mu)$ reaches unity gradually increases and reaches heights ranging between 2.7~Mm--5.7~Mm for the line center wavelength. 
However, as we can see from the source function ratio plots and the discussions below, the formation height of the linear polarization is different from that of the intensity.
To understand the formation of the $Q/I$ profiles we first define the wing region to be between [4225.4, 4226.59) and the core region to be [4226.59, 4226.75).
We then classify them according to their shapes in the core and the wings as shown in Figure~\ref{fig:classification}.
The $Q/I$ profiles are classified as those with two wing extrema (Figure~\ref{fig:classification}a), one wing extrema (Figure~\ref{fig:classification}b), two core extrema (Figure~\ref{fig:classification}c) and one core extrema (Figure~\ref{fig:classification}d). 
For convenience, the extremum closest to the continuum is classified as the primary extremum in the wing region, while in the core region, the line center extremum is designated as the primary extremum.
The cases with two extrema will then have an extremum we label as the secondary extremum.
In some cases, an extremum appears at the border wavelength between the wing and core region.
In these cases, depending on the wavelength value, we classify it as a secondary wing extremum or a secondary core extremum.
In the core region beside the the primary line center extremum we identify an extremum that we name the core minimum.
To understand the formation of polarization profiles we express the emergent $Q$ in terms of the source function term $\tilde{S}^2_0$ by substituting  Equation~(\ref{line-skq}) in Equation~(\ref{ikq-mup}) \citep[see expressions that relate $Q$ and $I^K_Q$ in Appendix B of][]{hf07}. The emergent $Q$ along the $\mu=0.3$ ray at a given $\lambda$ can be written as  

\begin{equation}
Q_{\lambda}(\tau_\lambda=0, \mu=0.3) \propto \int_0^{T_{\lambda}}
\exp^{-\tau^{\prime}_{\lambda}/{0.3}}  \tilde{S}^2_0 (\tau^{\prime}_{\lambda})\,\, d \tau^{\prime}_{\lambda} .
\label{contrib-q}
\end{equation}
The term inside the integral here is  the contribution function for $Q$.

\begin{figure*}[htbp]
    \centering
    \includegraphics[width=0.75\textwidth]{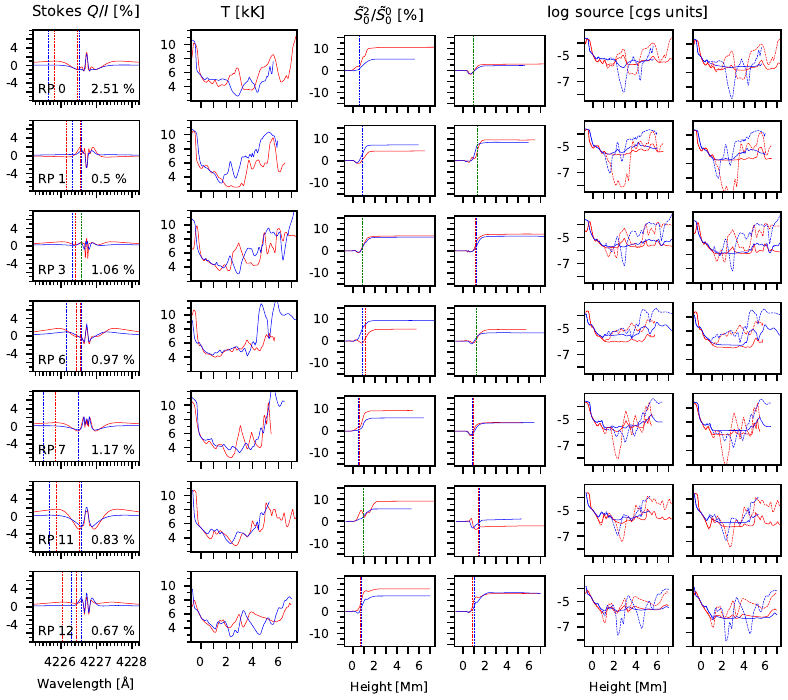}
    \caption{Synthesized Stokes~$Q/I$ profiles and other physical quantities extracted from the simulation for clusters with two wing extrema. Each row corresponds to one cluster indicated by the text in the first column. The first column presents the synthesized $Q/I$. The fraction of all profiles belonging to each cluster is indicated as a percentage next to the cluster number. The second column presents temperature stratification with height. The next two columns present the source function ratio as a function of height at wavelength positions marked by dashed vertical lines in order from shorter to longer wavelengths. 
    The vertical lines in the source function ratio plots indicate the height at which the $\log (\tau_{\lambda}/\mu)$ = 0, for the wavelength. The next two columns present the unpolarized total source function (solid) as a function of the height at the two wavelength positions, respectively. The Planck function (dashed) is overplotted for comparison. The vertical lines are red-colored for the red curve, blue-colored for the blue curve, and green if overlapping. The red and blue colors respectively indicate the profiles farthest and closest to the mean profile in each cluster.}
    \label{fig:hjbarwing1}
\end{figure*}

\begin{figure}[htbp]
    \centering
    \includegraphics[width=0.5\textwidth]{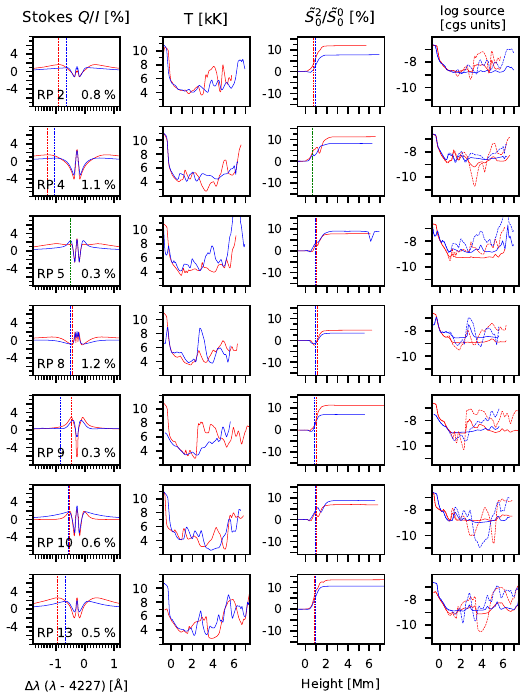}
    \caption{Synthesized Stokes~$Q/I$ profiles and other physical quantities extracted from the simulation for clusters with one wing extremum in the same format as Figure~\ref{fig:hjbarwing1}.}
    \label{fig:hjbarwing2}
\end{figure}

In the following sections we will interpret the emergent polarization signals by analyzing the source function ratios across the atmosphere, assuming that the whole height range where the source function ratio is non-zero, up to the height where it becomes constant, contributes to the formation of emergent $Q$ at a given wavelength $\lambda$ (refer to the Figures~\ref{fig:hjbarwing1}--\ref{fig:hjbarcore2} and the corresponding discussions). 
However, we remark here that the current analysis can be improved by more accurately determining the formation region of a given polarization feature using advanced techniques, such as response functions.

In Figures~\ref{fig:hjbarwing1}, \ref{fig:hjbarwing2}, \ref{fig:hjbarcore1} and \ref{fig:hjbarcore2} we plot height dependence of the temperature ($T$), wavelength dependence of the polarization profile ($Q/I$), height dependence of the source function ratio ($\tilde{S}^2_0/\tilde{S}^0_0$) and depth dependence of the unpolarized total source function in 14 chosen clusters.
The unpolarized total source function here comprises contributions from thermal and scattering components, along with both line and continuum contributions.
In these figures, each row corresponds to various quantities specific to a given cluster.
In each cluster, we show two different cases that represent profiles farthest (red) and closest (blue) to the mean profile in that cluster.

In Figures~\ref{fig:hjbarwing1} and \ref{fig:hjbarwing2}, we respectively show the cases with two extrema (peak or dip) and one extremum in the wing region.
Similarly, in Figures~\ref{fig:hjbarcore1} and Figure~\ref{fig:hjbarcore2}, we respectively show the profiles with two and one core extrema.
Vertical lines are drawn on the $Q/I$ profiles at selected wavelengths, for which extrema in the wings and core regions occur.
We also draw vertical lines at heights where $ \tau_{\lambda}/\mu  = 1 $ on the source function ratio plots.

In the last columns of each of these figures, we plot the height variation of the unpolarized total source function for the wavelength positions chosen for the source function ratio plots.
The Planck function for the respective wavelength points is also over-plotted.

%

\subsection{General behaviour} 
\label{sec:gb}

The atmospheres we consider have a maximum height of $\sim$ 7.5 Mm.
%
%
The temperature structure in these atmospheres shows steep gradients due to the magnetohydrodynamics present in the Bifrost simulation from which the 1D atmospheres are derived.

The $Q/I$ observations of the Ca {\sc i} 4227 \AA\, line show a typical triple peak structure.
However, for the atmospheres that we considered in this paper, the wavelength dependence of $Q/I$ shows a triple extrema structure, with peaks being positive or negative, for most of the rays.
This is because observations represent spatially and temporally averaged profiles while the MHD simulations of the atmosphere along each ray represent resolved regions on the Sun. 

The source function ratios that we plot behave very similar to the anisotropy factor plots shown in \citet[][]{rene2005}.
This is due to the coupling of the radiation field with the source function that includes PFR scattering.
Therefore in the following, we first discuss the general behavior of the anisotropy factor and then relate it to the behavior of the source function ratios.
As discussed in \citet{rene2005}, the sign and magnitude of the anisotropy and thus of the source function ratio are determined by the $\mu$-dependence of the radiation field.
The variation of the anisotropy across the atmosphere depends on the optical depth at the considered wavelength, and it is the result of a competition between the contributions from limb-darkened outgoing radiation and limb-brightened incoming radiation.
The limb-darkening of the outgoing radiation is due to the effect of the unpolarized total source function gradient, while the limb-brightening of the incoming radiation is due to the combination of the unpolarized total source function gradient and the boundary effects \citep[see also][]{2001ASPC..236..161T, 2004ASSL..307.....L,rene2005}.
In the deeper layers of the atmosphere, the contribution to the anisotropy from the outgoing limb-darkened radiation field cancels with that of the contribution from the incoming limb-brightened radiation field to give zero anisotropy.
In the higher layers, these effects compete with each other, and therefore, the anisotropy starts to become non-zero.
Initially, the anisotropy rises towards positive or negative values in the lower layer and reaches a constant, maximum value in the higher layer, where the radiation field decouples from local conditions.
The height at which the anisotropy reaches a constant value depends on the wavelength point.
This height increases as we go close to the line center, and for the line center wavelength, this constancy is reached very close to the top boundary.

Since the line core is formed in the atmosphere where the unpolarized total source function gradient is almost flat boundary effects control the anisotropy factor and also the source function ratio at this wavelength.

In the sections that follow, we explain the various spectral features seen in the wing and core region of the $Q/I$ profiles using the source function ratio  plots.

\subsection {Behaviour of $Q/I$ in the wing region}

The $Q/I$ profiles exhibit either one or two extrema in the wing region.
Among the chosen 14 clusters, 7 exhibit two wing extrema, and the other 7 exhibit one wing extremum.
These extrema are generated due to PFR effects.
Referring to the source function ratio plots at the specific wavelengths we can infer that $Q/I$ at the wing region originates between 0.5 Mm and 2 Mm in all the clusters.

In Figure~\ref{fig:hjbarwing1}, we show the profiles with two extrema in the line wing region.
Except RP 1 all other profiles have a positive line center peak.
Starting from the blue side continuum of the line center, the profiles have either a rising primary extremum followed by a falling secondary extremum (e.g., RP 0, RP 6, RP 7, and RP 11) or vice versa (e.g., RP 1, RP 3 and RP 12).

As seen from the source function ratio plots the primary wing extremum in polarization has a non-zero contribution of this ratio between 0.5~Mm to 1.5~Mm, at this wavelength (see 3rd column.).

For those cases which have positive primary extremum in the wings we can see that there is an accumulation of only positive contribution of source function ratio.
This creates a positive emergent linear polarization in the wings.
On the other hand, when there is a negative component of the source function ratio due to cancellation with the positive component, the emergent $Q/I$ value reduces and sometimes remains close to zero.

The secondary extremum in the polarization has the contribution of source function ratio between 0.5 Mm to 2 Mm (see 4th column).
The sign of the secondary extremum in the wing also has a correlation with the sign of the source function ratio at this wavelength.

We notice that the temperature gradients are monotonic, between 0.5~Mm to 2~Mm for most of the clusters, except RP 12, where we see the temperature decrease first and then increase.
We observe that when there is a negative temperature gradient with a cooler upper layer and hotter lower layer, we see a strong negative source function ratio in the lower layer that increases in the upper layer.
When the temperature gradient is positive, with a cooler lower layer and a hot upper layer, the source function ratio becomes positive in the lower layer and reaches a much higher positive value in the upper layer.

In Figure~\ref{fig:hjbarwing2}, we selected the profiles with only one extremum in the wings.
Strictly speaking, in some profiles in this category, we do have two extrema, but they do not appear as strong as the cases considered in Figure~\ref{fig:hjbarwing1}.
This extremum either appears in the far wings in the position around the first extremum in Figure~\ref{fig:hjbarwing1}  (RP 2, RP 4, RP 13) or it appears in the near wings, at the position of the second extremum in Figure~\ref{fig:hjbarwing1} (RP 5, RP 8, RP 10).

The case of RP 9 is an exception.
Because of our choice of wavelengths separating the core and the wing region, we have some ambiguity in selecting a particular extremum to be in the core or wing region.
To resolve this, we adhere to our definition of the core and wing regions, considering an extremum to be within the wing region only if it appears within that designated wavelength interval.
Therefore, we identify an extremum in the near wings for the red curve while the extremum is in the far wings for the blue curve, although both have similar shapes.

For all the cases considered in this figure, we observe that the temperature gradients are not as steep as those in Figure~\ref{fig:hjbarwing1} or have temperatures that decrease first and increase again in the region between 0.5~Mm to 2~Mm, eventually leading to non-monotonic temperature gradients.
The sign of the $Q/I$ extremum can be explained using source function ratio plots, employing arguments similar to those used for explaining Figure~\ref{fig:hjbarwing1}.

\subsection{Behaviour of $Q/I$ in the core region}
\begin{figure*}[htbp]
    \centering
    \includegraphics[width=\textwidth]{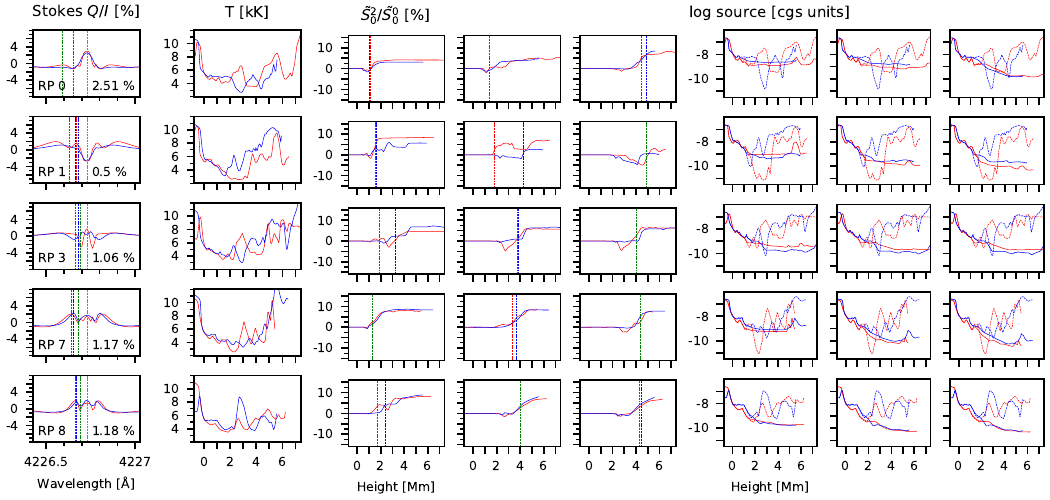}
    \caption{Synthesized Stokes~$Q/I$ profiles and other physical quantities extracted from the simulation for clusters with one secondary extremum and core minimum in the same format as Figure~\ref{fig:hjbarwing1}.}
    \label{fig:hjbarcore1}
\end{figure*}

\begin{figure*}[htbp]
    \centering
    \includegraphics[width=0.75\textwidth]{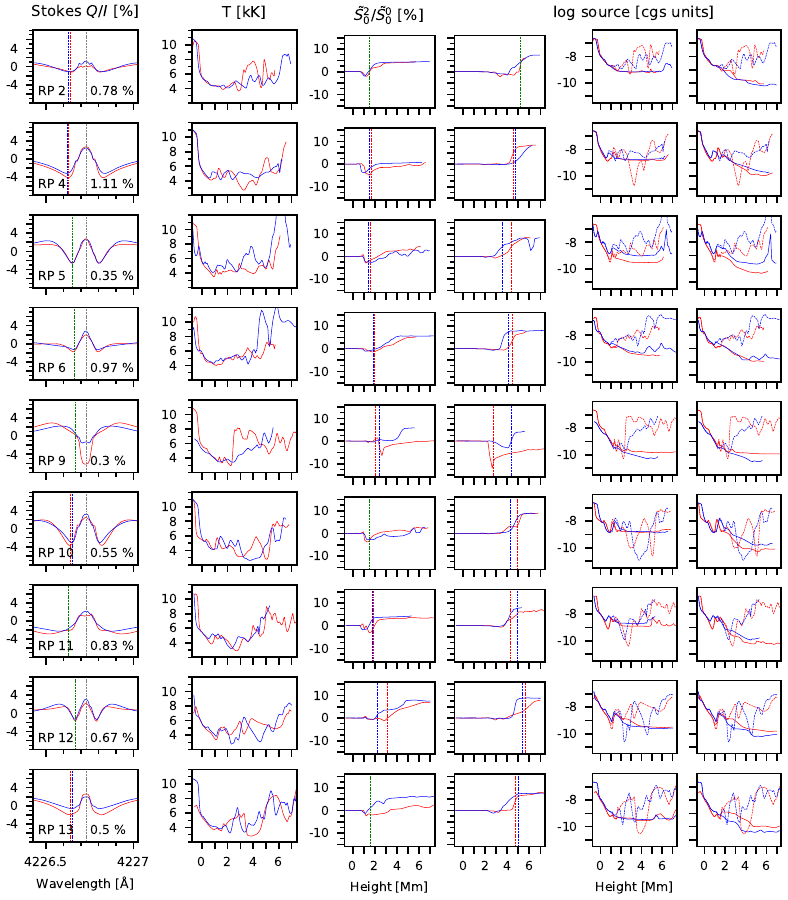}
    \caption{Synthesized Stokes~$Q/I$ profiles and other physical quantities extracted from the simulation for clusters with no secondary extremum but a core minimum in the same format as Figure~\ref{fig:hjbarwing1}.}
    \label{fig:hjbarcore2}
\end{figure*}

As we go closer to the line center, the complexity of the polarization profiles increases.
Typical features in the line core are a primary line center extremum, a  secondary extrema, and a core minimum, defined as the smallest non-zero absolute value of the polarization.
We classify profiles with respect to the presence of a secondary extremum (plotted in Figure~\ref{fig:hjbarcore1}) or its absence (plotted in Figure~\ref{fig:hjbarcore2}) in the core region close to the line center.
In the following, we explain two polarization features in the core region, namely, the secondary extremum when present and the core minimum.
The line center extremum is discussed in a separate subsection.

The secondary core extremum appears in various ways: same sign as the line center polarization but a smaller absolute magnitude, opposite sign as the line center polarization but a smaller absolute magnitude, same sign as the line center polarization but with a higher absolute magnitude.

From depth dependence of source function ratio and that of the unpolarized total source function at the secondary core extremum's wavelength, we can find that this feature forms at a range of heights between 2~Mm and 4~Mm.

When formed around 2~Mm, it is farther from the line center when compared to its proximity when formed around 4 Mm.
For example, for RP 0, source function ratio contribution is between 0.5~Mm to 2~Mm, while it is between 1~Mm to 2.5~Mm in the case of RP 7.
As the contribution from higher layers increases, the source function ratio also increases.
Therefore, for example, the case of RP 0 has a lower value of source function ratio in the secondary core extremum than that at the line center, while they are comparable in the case of RP 7 (compare third and fifth columns of Figure~\ref{fig:hjbarcore1}). 

Source function ratio plots in Figure~\ref{fig:hjbarcore2} reach a constant maximum value in the higher layers, which, in general, is much lower than those in Figure~\ref{fig:hjbarcore1}.
Because of this, the rise from lower to a constant higher value in source function ratio appears less steep in Figure~\ref{fig:hjbarcore2} when compared to those in Figure~\ref{fig:hjbarcore1}.
This higher magnitude of source function ratio in Figure~\ref{fig:hjbarcore1} correlates with the presence of secondary extremum in $Q/I$ in that Figure.
Also, the unpolarized total source function plots suggest that the profiles with secondary core extremum have steeper gradients.

In all the clusters we considered, the primary extremum (at line center) and secondary core extremum are either both maxima or both minima.
In between these two extrema is the extremum, which we refer to as the core minimum with opposite orientation. 

For some clusters, the core minimum does not appear as an extremum but simply represents the wavelength at which polarization in absolute values reaches minimum value in the defined wavelength domain for the core region (see, e.g., RP 9 and RP 11).
In such a case polarization transition from line center to wing is monotonic.
For this cluster, at the core, minimum wavelength polarization is nearly zero, as both the positive and negative source function ratio contributions cancel. 

In most clusters, polarization at the core minimum wavelength is negative because of the dominant negative source function ratio contribution between 1.1~Mm and 3.1~Mm.

\subsection{Behaviour of line center polarization}

In Figure~\ref{fig:tau}, $\log (\tau_{\lambda}/\mu)$ is plotted for various clusters.
We can see here that
the line center intensity is formed in the range of 2.7~Mm--5.7~Mm.

Source function ratio at the line center is zero until 2.5~Mm and contributes to line center polarization between 2.5~Mm to the top of the atmosphere.
Line center extremum in polarization either appears as a positive peak or as a negative dip.
The sign of $Q/I$ at the line center is positive (negative) if positive (negative) source function ratio dominates above 2.5~Mm.

The last panel plots the unpolarized total source function as a function of height.
The overplotted Planck functions clearly show that the radiation field decouples from the local conditions at around 2~Mm. The fluctuations of the Planck function are smooth compared to those of the source function for all the cases.
The unpolarized total source function at the line center wavelength becomes almost flat in the regions where line core is formed, causing the boundary effects to dominate \citep[see discussions in Section~\ref{sec:gb}
 and also][]{rene2005}.

\section{Conclusions}
\label{sect:conclusions}

In this paper, we present an analysis of the formation of the $Q/I$ profiles of the Ca {\sc i} 4227 \AA\, line, which are formed in various atmospheres extracted from a publicly available Bifrost simulation.
Our aim in this paper is to understand the differences in the formation of the polarization profiles in simulated resolved atmospheres in comparison to 1D semi-empirical atmospheres.
We have considered the effects of the temperature stratification and the source function ratio on the polarization profiles.
The source function ratio behaves qualitatively very similar to the anisotropy factor discussed in \citep[][]{rene2005}.

We found that the $Q/I$ profiles from the atmospheres extracted from the simulation show a ``triple extrema structure'' instead of the observed ``triple peak structure''.
We explain the emergent $Q/I$ profiles using the depth dependence of the source function ratio at various wavelength positions.
In the case of line wings, the source function ratio is correlated with the temperature gradients at the heights of polarization formation.
However, in the line core, the source function ratio cannot be explained using temperature gradients because of the non-LTE RT effects.

As stated earlier, we have analyzed the effect of resonance scattering on the polarization profiles by studying the source function ratios.  The velocity and magnetic fields were not included.
The Hanle effect involves the modification of these polarization profiles in the presence of magnetic fields with the generation of a non-zero $U/I$.
Since the Hanle effect only applies in the core of the line,  we expect that the field strengths affect the magnitude of the Stokes~$Q/I$ profiles while the orientation of magnetic fields will affect the shapes of $Q/I,U/I$ profiles in the line core.
Moreover, the magneto-optical effects can alter the polarization in the line wings. 
On the other hand, the velocity fields introduce asymmetries in both the intensity and polarization profiles and thus, may result in a significant change in the Stokes~$Q/I$ profiles.

This study presents a comprehensive analysis of synthetic $Q/I$ profiles of the Ca {\sc i} 4227 \AA\, line generated from a realistic quiet-Sun simulation.
By elucidating the intricate relationships between temperature structure and source function ratio with the emergent linear polarization profiles, this research provides essential insights necessary for analyzing and interpreting the unprecedentedly detailed data that next-generation telescopes like DKIST will generate.

\begin{acknowledgments}
The authors thank the anonymous reviewer whose feedback greatly improved the manuscript.
This research has made use of the High-Performance Computing (HPC) resources (NOVA cluster) made available by the Computer Center of the Indian Institute of Astrophysics, Bangalore.
LSA would like to thank Prof. H. Frisch for useful discussions.

\end{acknowledgments}

\vspace{5mm}

\software{NumPy\citep{2020Natur.585..357H},
          matplotlib\citep{2007CSE.....9...90H},
          RH\citep{2001ApJ...557..389U},
          POLY\citep[][]{rene2005,LSA2011}
          }

\bibliography{sample631}{}

\begin{thebibliography}{}
\expandafter\ifx\csname natexlab\endcsname\relax\def\natexlab#1{#1}\fi
\providecommand{\url}[1]{\href{#1}{#1}}
\providecommand{\dodoi}[1]{doi:~\href{http://doi.org/#1}{\nolinkurl{#1}}}
\providecommand{\doeprint}[1]{\href{http://ascl.net/#1}{\nolinkurl{http://ascl.net/#1}}}
\providecommand{\doarXiv}[1]{\href{https://arxiv.org/abs/#1}{\nolinkurl{https://arxiv.org/abs/#1}}}

\bibitem[{{Alsina Ballester} {et~al.}(2018){Alsina Ballester}, {Belluzzi}, \& {Trujillo Bueno}}]{2018ApJ...854..150A}
{Alsina Ballester}, E., {Belluzzi}, L., \& {Trujillo Bueno}, J. 2018, \apj, 854, 150, \dodoi{10.3847/1538-4357/aa978a}

\bibitem[{{Anusha} \& {Nagendra}(2011)}]{LSA3D2011}
{Anusha}, L.~S., \& {Nagendra}, K.~N. 2011, \apj, 726, 6, \dodoi{10.1088/0004-637X/726/1/6}

\bibitem[{{Anusha} {et~al.}(2021){Anusha}, {Shapiro}, {Witzke}, {Cernetic}, {Solanki}, \& {Gizon}}]{2021ApJS..255....3A}
{Anusha}, L.~S., {Shapiro}, A.~I., {Witzke}, V., {et~al.} 2021, \apjs, 255, 3, \dodoi{10.3847/1538-4365/abfb72}

\bibitem[{{Anusha} {et~al.}(2011){Anusha}, {Nagendra}, {Bianda}, {Stenflo}, {Holzreuter}, {Sampoorna}, {Frisch}, {Ramelli}, \& {Smitha}}]{LSA2011}
{Anusha}, L.~S., {Nagendra}, K.~N., {Bianda}, M., {et~al.} 2011, \apj, 737, 95, \dodoi{10.1088/0004-637X/737/2/95}

\bibitem[{{Avrett}(1985)}]{1985cdm..proc...67A}
{Avrett}, E.~H. 1985, in Chromospheric Diagnostics and Modelling, ed. B.~W. {Lites}, 67--127

\bibitem[{{Bianda} {et~al.}(2011){Bianda}, {Ramelli}, {Anusha}, {Stenflo}, {Nagendra}, {Holzreuter}, {Sampoorna}, {Frisch}, \& {Smitha}}]{2011A&A...530L..13B}
{Bianda}, M., {Ramelli}, R., {Anusha}, L.~S., {et~al.} 2011, \aap, 530, L13, \dodoi{10.1051/0004-6361/201117047}

\bibitem[{{Bommier}(1997{\natexlab{a}})}]{vb97a}
{Bommier}, V. 1997{\natexlab{a}}, \aap, 328, 706

\bibitem[{{Bommier}(1997{\natexlab{b}})}]{vb97b}
---. 1997{\natexlab{b}}, \aap, 328, 726

\bibitem[{{Capozzi} {et~al.}(2020){Capozzi}, {Alsina Ballester}, {Belluzzi}, {Bianda}, {Dhara}, \& {Ramelli}}]{2020A&A...641A..63C}
{Capozzi}, E., {Alsina Ballester}, E., {Belluzzi}, L., {et~al.} 2020, \aap, 641, A63, \dodoi{10.1051/0004-6361/202038455}

\bibitem[{{Carlin} \& {Bianda}(2016)}]{2016ApJ...831L...5C}
{Carlin}, E.~S., \& {Bianda}, M. 2016, \apjl, 831, L5, \dodoi{10.3847/2041-8205/831/1/L5}

\bibitem[{{Carlin} \& {Bianda}(2017)}]{2017ApJ...843...64C}
---. 2017, \apj, 843, 64, \dodoi{10.3847/1538-4357/aa7800}

\bibitem[{{Carlsson} {et~al.}(2016){Carlsson}, {Hansteen}, {Gudiksen}, {Leenaarts}, \& {De Pontieu}}]{2016A&A...585A...4C}
{Carlsson}, M., {Hansteen}, V.~H., {Gudiksen}, B.~V., {Leenaarts}, J., \& {De Pontieu}, B. 2016, \aap, 585, A4, \dodoi{10.1051/0004-6361/201527226}

\bibitem[{{Faurobert-Scholl}(1992)}]{mf92}
{Faurobert-Scholl}, M. 1992, \aap, 258, 521

\bibitem[{{Fontenla} {et~al.}(1993{\natexlab{a}}){Fontenla}, {Avrett}, \& {Loeser}}]{1993ApJ...406..319F}
{Fontenla}, J.~M., {Avrett}, E.~H., \& {Loeser}, R. 1993{\natexlab{a}}, \apj, 406, 319, \dodoi{10.1086/172443}

\bibitem[{{Fontenla} {et~al.}(1993{\natexlab{b}}){Fontenla}, {Avrett}, \& {Loeser}}]{fontenlaetal93}
---. 1993{\natexlab{b}}, \apj, 406, 319, \dodoi{10.1086/172443}

\bibitem[{{Frisch}(2007)}]{hf07}
{Frisch}, H. 2007, \aap, 476, 665, \dodoi{10.1051/0004-6361:20077980}

\bibitem[{{Frisch}(2022)}]{hf2022}
---. 2022, {Radiative Transfer: An Introduction to Exact and Asymptotic Methods}, \dodoi{10.1007/978-3-030-95247-1}

\bibitem[{{Gudiksen} {et~al.}(2011){Gudiksen}, {Carlsson}, {Hansteen}, {Hayek}, {Leenaarts}, \& {Mart{\'\i}nez-Sykora}}]{2011A&A...531A.154G}
{Gudiksen}, B.~V., {Carlsson}, M., {Hansteen}, V.~H., {et~al.} 2011, \aap, 531, A154, \dodoi{10.1051/0004-6361/201116520}

\bibitem[{{Guerreiro} {et~al.}(2024){Guerreiro}, {Janett}, {Riva}, {Benedusi}, \& {Belluzzi}}]{2024A&A...683A.207G}
{Guerreiro}, N., {Janett}, G., {Riva}, S., {Benedusi}, P., \& {Belluzzi}, L. 2024, \aap, 683, A207, \dodoi{10.1051/0004-6361/202346399}

\bibitem[{{Harris} {et~al.}(2020){Harris}, {Millman}, {van der Walt}, {Gommers}, {Virtanen}, {Cournapeau}, {Wieser}, {Taylor}, {Berg}, {Smith}, {Kern}, {Picus}, {Hoyer}, {van Kerkwijk}, {Brett}, {Haldane}, {del R{\'\i}o}, {Wiebe}, {Peterson}, {G{\'e}rard-Marchant}, {Sheppard}, {Reddy}, {Weckesser}, {Abbasi}, {Gohlke}, \& {Oliphant}}]{2020Natur.585..357H}
{Harris}, C.~R., {Millman}, K.~J., {van der Walt}, S.~J., {et~al.} 2020, \nat, 585, 357, \dodoi{10.1038/s41586-020-2649-2}

\bibitem[{{Holzreuter} {et~al.}(2005){Holzreuter}, {Fluri}, \& {Stenflo}}]{rene2005}
{Holzreuter}, R., {Fluri}, D.~M., \& {Stenflo}, J.~O. 2005, \aap, 434, 713, \dodoi{10.1051/0004-6361:20042096}

\bibitem[{{Hunter}(2007)}]{2007CSE.....9...90H}
{Hunter}, J.~D. 2007, Computing in Science and Engineering, 9, 90, \dodoi{10.1109/MCSE.2007.55}

\bibitem[{{Jaume Bestard} {et~al.}(2021){Jaume Bestard}, {Trujillo Bueno}, {{\v{S}}t{\v{e}}p{\'a}n}, \& {del Pino Alem{\'a}n}}]{2021ApJ...909..183J}
{Jaume Bestard}, J., {Trujillo Bueno}, J., {{\v{S}}t{\v{e}}p{\'a}n}, J., \& {del Pino Alem{\'a}n}, T. 2021, \apj, 909, 183, \dodoi{10.3847/1538-4357/abd94a}

\bibitem[{{Joshi} \& {Rouppe van der Voort}(2022)}]{2022A&A...664A..72J}
{Joshi}, J., \& {Rouppe van der Voort}, L. H.~M. 2022, \aap, 664, A72, \dodoi{10.1051/0004-6361/202243051}

\bibitem[{{Landi Degl'Innocenti} \& {Landolfi}(2004{\natexlab{a}})}]{ll04}
{Landi Degl'Innocenti}, E., \& {Landolfi}, M. 2004{\natexlab{a}}, {Polarization in Spectral Lines}, Vol. 307, \dodoi{10.1007/978-1-4020-2415-3}

\bibitem[{{Landi Degl'Innocenti} \& {Landolfi}(2004{\natexlab{b}})}]{2004ASSL..307.....L}
---. 2004{\natexlab{b}}, {Polarization in Spectral Lines}, Vol. 307, \dodoi{10.1007/978-1-4020-2415-3}

\bibitem[{MacQueen(1967)}]{zbMATH03340881}
MacQueen, J. 1967, Some methods for classification and analysis of multivariate observations, Proc. 5th {Berkeley} {Symp}. {Math}. {Stat}. {Probab}., {Univ}. {Calif}. 1965/66, 1, 281-297 (1967).

\bibitem[{{Mathur} {et~al.}(2022){Mathur}, {Joshi}, {Nagaraju}, {Rouppe van der Voort}, \& {Bose}}]{2022A&A...668A.153M}
{Mathur}, H., {Joshi}, J., {Nagaraju}, K., {Rouppe van der Voort}, L., \& {Bose}, S. 2022, \aap, 668, A153, \dodoi{10.1051/0004-6361/202244332}

\bibitem[{{Moe} {et~al.}(2023){Moe}, {Pereira}, {Calvo}, \& {Leenaarts}}]{2023A&A...675A.130M}
{Moe}, T.~E., {Pereira}, T. M.~D., {Calvo}, F., \& {Leenaarts}, J. 2023, \aap, 675, A130, \dodoi{10.1051/0004-6361/202346724}

\bibitem[{{Moe} {et~al.}(2024){Moe}, {Pereira}, {Rouppe van der Voort}, {Carlsson}, {Hansteen}, {Calvo}, \& {Leenaarts}}]{2024A&A...682A..11M}
{Moe}, T.~E., {Pereira}, T. M.~D., {Rouppe van der Voort}, L., {et~al.} 2024, \aap, 682, A11, \dodoi{10.1051/0004-6361/202347328}

\bibitem[{{Panos} {et~al.}(2018){Panos}, {Kleint}, {Huwyler}, {Krucker}, {Melchior}, {Ullmann}, \& {Voloshynovskiy}}]{2018ApJ...861...62P}
{Panos}, B., {Kleint}, L., {Huwyler}, C., {et~al.} 2018, \apj, 861, 62, \dodoi{10.3847/1538-4357/aac779}

\bibitem[{{Przybylski} {et~al.}(2022){Przybylski}, {Cameron}, {Solanki}, {Rempel}, {Leenaarts}, {Anusha}, {Witzke}, \& {Shapiro}}]{2022A&A...664A..91P}
{Przybylski}, D., {Cameron}, R., {Solanki}, S.~K., {et~al.} 2022, \aap, 664, A91, \dodoi{10.1051/0004-6361/202141230}

\bibitem[{{Rimmele} {et~al.}(2020){Rimmele}, {Warner}, {Keil}, {Goode}, {Kn{\"o}lker}, {Kuhn}, {Rosner}, {McMullin}, {Casini}, {Lin}, {W{\"o}ger}, {von der L{\"u}he}, {Tritschler}, {Davey}, {de Wijn}, {Elmore}, {Fehlmann}, {Harrington}, {Jaeggli}, {Rast}, {Schad}, {Schmidt}, {Mathioudakis}, {Mickey}, {Anan}, {Beck}, {Marshall}, {Jeffers}, {Oschmann}, {Beard}, {Berst}, {Cowan}, {Craig}, {Cross}, {Cummings}, {Donnelly}, {de Vanssay}, {Eigenbrot}, {Ferayorni}, {Foster}, {Galapon}, {Gedrites}, {Gonzales}, {Goodrich}, {Gregory}, {Guzman}, {Guzzo}, {Hegwer}, {Hubbard}, {Hubbard}, {Johansson}, {Johnson}, {Liang}, {Liang}, {McQuillen}, {Mayer}, {Newman}, {Onodera}, {Phelps}, {Puentes}, {Richards}, {Rimmele}, {Sekulic}, {Shimko}, {Simison}, {Smith}, {Starman}, {Sueoka}, {Summers}, {Szabo}, {Szabo}, {Wampler}, {Williams}, \& {White}}]{2020SoPh..295..172R}
{Rimmele}, T.~R., {Warner}, M., {Keil}, S.~L., {et~al.} 2020, SoPh, 295, 172, \dodoi{10.1007/s11207-020-01736-7}

\bibitem[{{Stein} \& {Nordlund}(1998)}]{1998ApJ...499..914S}
{Stein}, R.~F., \& {Nordlund}, {\r{A}}. 1998, \apj, 499, 914, \dodoi{10.1086/305678}

\bibitem[{{Supriya} {et~al.}(2014){Supriya}, {Smitha}, {Nagendra}, {Stenflo}, {Bianda}, {Ramelli}, {Ravindra}, \& {Anusha}}]{2014ApJ...793...42S}
{Supriya}, H.~D., {Smitha}, H.~N., {Nagendra}, K.~N., {et~al.} 2014, \apj, 793, 42, \dodoi{10.1088/0004-637X/793/1/42}

\bibitem[{{Trujillo Bueno}(2001)}]{2001ASPC..236..161T}
{Trujillo Bueno}, J. 2001, in Astronomical Society of the Pacific Conference Series, Vol. 236, Advanced Solar Polarimetry -- Theory, Observation, and Instrumentation, ed. M.~{Sigwarth}, 161, \dodoi{10.48550/arXiv.astro-ph/0202328}

\bibitem[{{Uitenbroek}(2001)}]{2001ApJ...557..389U}
{Uitenbroek}, H. 2001, \apj, 557, 389, \dodoi{10.1086/321659}

\bibitem[{{Vernazza} {et~al.}(1981){Vernazza}, {Avrett}, \& {Loeser}}]{val81}
{Vernazza}, J.~E., {Avrett}, E.~H., \& {Loeser}, R. 1981, \apjs, 45, 635, \dodoi{10.1086/190731}

\bibitem[{{V{\"o}gler} {et~al.}(2005){V{\"o}gler}, {Shelyag}, {Sch{\"u}ssler}, {Cattaneo}, {Emonet}, \& {Linde}}]{2005A&A...429..335V}
{V{\"o}gler}, A., {Shelyag}, S., {Sch{\"u}ssler}, M., {et~al.} 2005, \aap, 429, 335, \dodoi{10.1051/0004-6361:20041507}

\end{thebibliography}
\bibliographystyle{aasjournal}

\appendix

\section{$k$-means clustering}
\label{sect:appkmeans}

\begin{figure}
    \centering
    \includegraphics[width=0.5\textwidth]{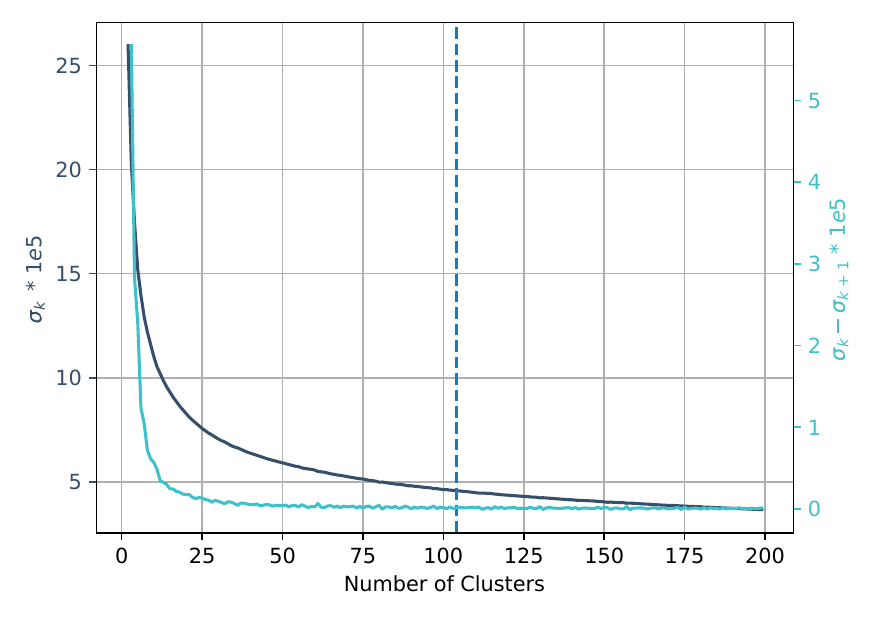}
    \caption{Determining the optimum number of clusters $k$ for the $k$-means clustering of $Q/I$ profiles of the Ca {\sc i} 4227 \AA\, line. The dark blue curve represents inertia ($\sigma_{k}$) for k clusters, while the cyan curve shows $\sigma_{k} - \sigma_{k+1}$. The dashed vertical line indicates the selected number of clusters $k=105$.}
    \label{fig:inertia}
\end{figure}
The $k$-means clustering algorithm \citep{zbMATH03340881} is a widely used unsupervised learning method for identifying patterns and structures within an unlabeled data set.
The algorithm partitions an unlabeled data set, consisting of $m$ data points with $n$ features, into $k$ clusters.
In our study, the unlabeled data set comprises the 15876 pixels selected for the forward modeling, and the features are 299 wavelength positions.
The line is sampled within the range of $\pm$1.5~\AA\, on either side of the line core, with a spectral sampling of 0.01~\AA.
The $k$-means algorithm is iterative and aims to minimize inertia ($\sigma_{k}$), which is the within-cluster sum of squared distances from the cluster center. A cluster center represents the mean of all data points within that cluster.
Specifically, in our study, a cluster center is the mean of all spectral profiles within the cluster, referred to as the RP.
%

Our primary aim in grouping the data was to identify uniquely shaped $Q/I$ profiles.
Ideally, all profiles within a cluster should closely resemble the mean profile, RP, of that cluster.
The $k$-means algorithm calculates the Euclidean distance between a data point and the cluster mean, implying that higher variance wavelength positions affect the $\chi^{2}$ more significantly.
Therefore, we normalized the data by subtracting the mean and dividing it by the standard deviation at each wavelength position before clustering, ensuring all positions had a variance of one.
From the literature and a preliminary look at the data, we understood that there was a huge difference in amplitude with respect to the wavelength of the $Q/I$ profiles.
Moreover, the number of wavelength positions in the far wing, $\Delta \lambda < -$0.5~\AA\, and $\Delta \lambda > +$0.5~\AA\,(199 wavelength positions), were a lot more than in the core region, $\Delta \lambda \in \pm$0.5~\AA\, (100 wavelength positions), of the Ca {\sc i} 4227 \AA\, line.
Further, we have also defined an inner core region with 50 wavelength positions, $\Delta \lambda \in \pm$0.25~\AA.
To ensure a clear distinction among clusters with various shapes and to give equal priority to all wavelength regions, we applied weights to the data points.
These weights were adjusted so that the sum of variances of wavelength positions in the far wing region matched the sum of variances in the inner core region, and both were equal to the sum of variances in the outer core region.

The elbow method is employed to identify the optimal number of clusters, $k$.
A relatively large value, $k=105$, is selected to clearly distinguish between different shapes of $Q/I$ profiles, following the approach outlined in \citet{2022A&A...668A.153M};
see also Figure~\ref{fig:inertia}.

%
%
%
%

\section{Characterization of RPs}\label{sect:charac}

As demonstrated in Appendix~\ref{sect:appkmeans}, the number of clusters initially used was significantly higher than necessary.
We have reclassified the remaining 91 clusters into one of the original 14 clusters.
In Table~\ref{tab:rpreclassify}, the first column lists the indices of the first 14 clusters [0–13], while the second column indicates which of the remaining clusters exhibit similar shapes to each of these clusters.
Cluster 0 has the highest number of similar clusters, with 16 clusters having Stokes Q RP shapes comparable to the RP of Cluster 0.
Most other clusters are similar to 4 or 5 other clusters, with Clusters 10 and 13 having the fewest similarities, each corresponding to only 3 clusters.

As discussed in the main text, the RPs are classified based on two criteria.
The first criterion is whether the RPs have one or two extrema in the wings of the polarization profiles.
RPs 0, 1, 3, 6, 7, 11, and 12, along with their similar RPs, exhibit two extrema in the line wings, while RPs 2, 4, 5, 8, 9, 10, and 13, and their similar RPs, have one extremum in the line wings.
The second criterion considers the presence of a secondary extremum near the line core of the polarization profiles.
RPs 0, 1, 3, 7, and 8, and their similar RPs, display one secondary extremum and a core minimum, while RPs 2, 4, 5, 6, 9, 10, 11, 12, and 13 have only a core minimum.

\begin{table}[htbp]
    \centering
    \begin{tabular}{cc}
    \hline\hline
    Initial cluster [0 -- 13] & Similar clusters\\
    \hline
    0 & 20, 21, 22, 23, 24, 31, 32, 55, 58, 60, 75, 81, 83, 87, 88, 95\\
    1 & 16, 37, 67, 92, 102\\
    2 & 18, 28, 51, 59, 85, 89, 93\\
    3 & 30, 41, 50, 65, 94, 97\\
    4 & 33, 34, 45, 56, 69, 74, 91, 99, 100, 104\\
    5 & 19, 49, 76, 101\\
    6 & 15, 42, 44, 48, 68, 73, 79, 80\\
    7 & 26, 35, 62, 64, 90\\
    8 & 25, 27, 47, 72, 77, 96, 103\\
    9 & 14, 38, 70, 71, 86\\
    10 & 17, 52, 53\\
    11 & 36, 46, 57, 66, 84, 98\\
    12 & 29, 39, 40, 43, 54, 82\\
    13 & 61, 63, 78\\
    \hline
    \end{tabular}
    \caption{Reclassification of clusters based on shape similarities. The first column lists the indices of the original 14 clusters [0--13]. The second column shows which of the remaining 91 clusters have polarization profiles similar to each of these 14 clusters.}
    \label{tab:rpreclassify}
\end{table}

\end{document}